\documentclass[prl,twocolumn, superscriptaddress, a4paper, showpacs]{revtex4}
\usepackage{amssymb,amsfonts,amsmath}
\usepackage{graphicx}

\begin{document}
\title{Realization of L\'evy flights as continuous processes}
\author{Ihor Lubashevsky}
    \affiliation{A.M. Prokhorov General Physics Institute, Russian
    Academy of Sciences, Vavilov Str. 38, 119991 Moscow, Russia}
    \affiliation{Institut f\"ur Physikalische Chemie,
    Westf\"alische Wilhelms Universit\"at M\"unster,  Corrensstr. 30,
    48149 M\"unster, Germany}
\author{Rudolf Friedrich}
    \affiliation{Institut f\"ur Theoretische Physik,
    Westf\"alische Wilhelms Universit\"at M\"unster, Wilhelm-Klemm. 9, 48149 M\"unster,
    Germany}
    \affiliation{Center of Nonlinear Science CeNoS,
    Westf\"alische Wilhelms Universit\"at M\"unster, 48149 M\"unster, Germany}
\author{Andreas Heuer}
    \affiliation{Institut f\"ur Physikalische Chemie, Westf\"alische Wilhelms
    Universit\"at M\"unster,  Corrensstr. 30,
    48149 M\"unster, Germany}
    \affiliation{Center of Nonlinear Science CeNoS,
    Westf\"alische Wilhelms Universit\"at M\"unster, 48149 M\"unster, Germany}
\date{\today}
\begin{abstract}
    On the basis of multivariate Langevin processes
    we present a realization of L\'evy flights as a continuous process.
    For the simple case of a particle moving under the influence
    of friction and a velocity dependent stochastic force we explicitly
    derive the generalized Langevin equation and the corresponding
    generalized Fokker-Planck equation describing L\'evy flights.
    Our procedure is similar to the treatment of the Kramers-Fokker Planck
    equation in the Smoluchowski limit.
    The proposed approach forms a feasible way of tackling L\'evy flights in
    inhomogeneous media or systems with boundaries what is up to now a
    challenging problem.
\end{abstract}


\maketitle

It has become evident that Brownian random walks may be a too simple
description of diffusion processes in complex systems like the motion of tracer
particles in turbulent flows \cite{Swinney}, the diffusion of particles in
random media \cite{Bouchaud}, the motion of wandering albatrosses \cite{Vis},
human travel behavior and spreading of epidemics \cite{Brockmann} or economic
time series in finance \cite{Stanley}. A variable $x$ corresponding to such a
process can frequently exhibit the dynamics described by the notion of
superfast diffusion, where the characteristic value $\bar{x}$ of the variable
$x$ demonstrates scaling behavior $[\bar{x}(t)]^{2} \propto t^{2/\alpha}$ with
$\alpha < 2$ \cite{Physicstoday,Lect}.

Brownian motion is described on the basis of Langevin equations or, in a
statistical sense, by the Fokker-Planck equation (cf.
\cite{LRisken,LGardiner}). A straightforward way to deal with anomalous
diffusion is based on a generalization of the Langevin equations by replacing
Gaussian white noise with L\'evy noise \cite{Fogedby1}. Recently, there has
been a great deal of research about superfast diffusion. It includes, in
particular, a rather general analysis of the Langevin equation with L\'evy
noise (see, e.g., Ref.~\cite{Weron}) and the form of the corresponding
Fokker-Planck equations \cite{Schertzer1,Schertzer2}, description of anomalous
diffusion with power law distributions of spatial and temporal steps
\cite{Fogedby1,Sokolov}, L\'evy flights in heterogeneous media
\cite{Fogedby2,Honkonen,BrockmannGeisel} and in external fields
\cite{BrockmannSokolov,Fogedby3}, first passage time analysis and escaping
problem for L\'evy flights \cite{fptp1,fptp2,fptp3,fptp4,fptp5,fptp6}, as well
as processing experimental data for detecting the L\'evy type
behavior~\cite{SiegertLevy}. Besides, it should be noted that the attempt to
consider L\'evy flights in bounded systems (see, e.g., Ref.~\cite{nmlf1,nmlf2}
and review~\cite{nmlf3} as well) has introduced the notion of L\'evy walks
being a non-Markovian process because of the necessity to bound the walker
velocity.

The key point in constructing the mutually related pair of the stochastic
Langevin equation and the nonlocal Fokker-Planck equation for superdiffusion is
the L\'evy-Gnedenko central limit theorem
\cite{Schertzer1,Schertzer2,L'evy-Gned}. For the superdiffusion processes it
specifies the possible step distributions $P(\Delta x)$ which are universal and
actually independent of the details in the system behavior at the microscopic
level. In particular, for a symmetrical homogeneous 1D system superdiffusion
can be regarded as a chain of steps $\{\Delta x\}$ of duration $\delta t$ whose
distribution function $P(\Delta x)$ exhibits the following asymptotic behavior
for $|\Delta x|\gg \bar{x}(\delta t)$
\begin{equation}\label{LevyKernel}
    P(\Delta x)\sim \frac{[\bar{x}(\delta t)]^\alpha}{|\Delta x|^{\alpha+1}}\,.
\end{equation}
In spite of the considerable success achieved in this field the
theory of superdiffusion is far from being completed. For a given
elementary step of any small duration it is impossible to single out
some bounded domain that contains its initial $x_i$ and terminal
$x_t$ points with the probability practically equal to unity because
the second moment $\left<(x_t-x_i)^2\right>$ diverges. This renders
the description of L\'evy flights in heterogeneous media or media
with boundaries a challenging problem. Within the classical
formulation the L\'evy flight is not a spatially continuous
processes. As a consequence it is not possible to attribute local
characteristics to L\'evy flights which might help to identify,
e.g., the encounter time with boundaries. Bounding the particle
velocity breaks the L\'evy as well as the Markov properties.

\begin{figure}
\begin{center}
\includegraphics[width=0.9\columnwidth]{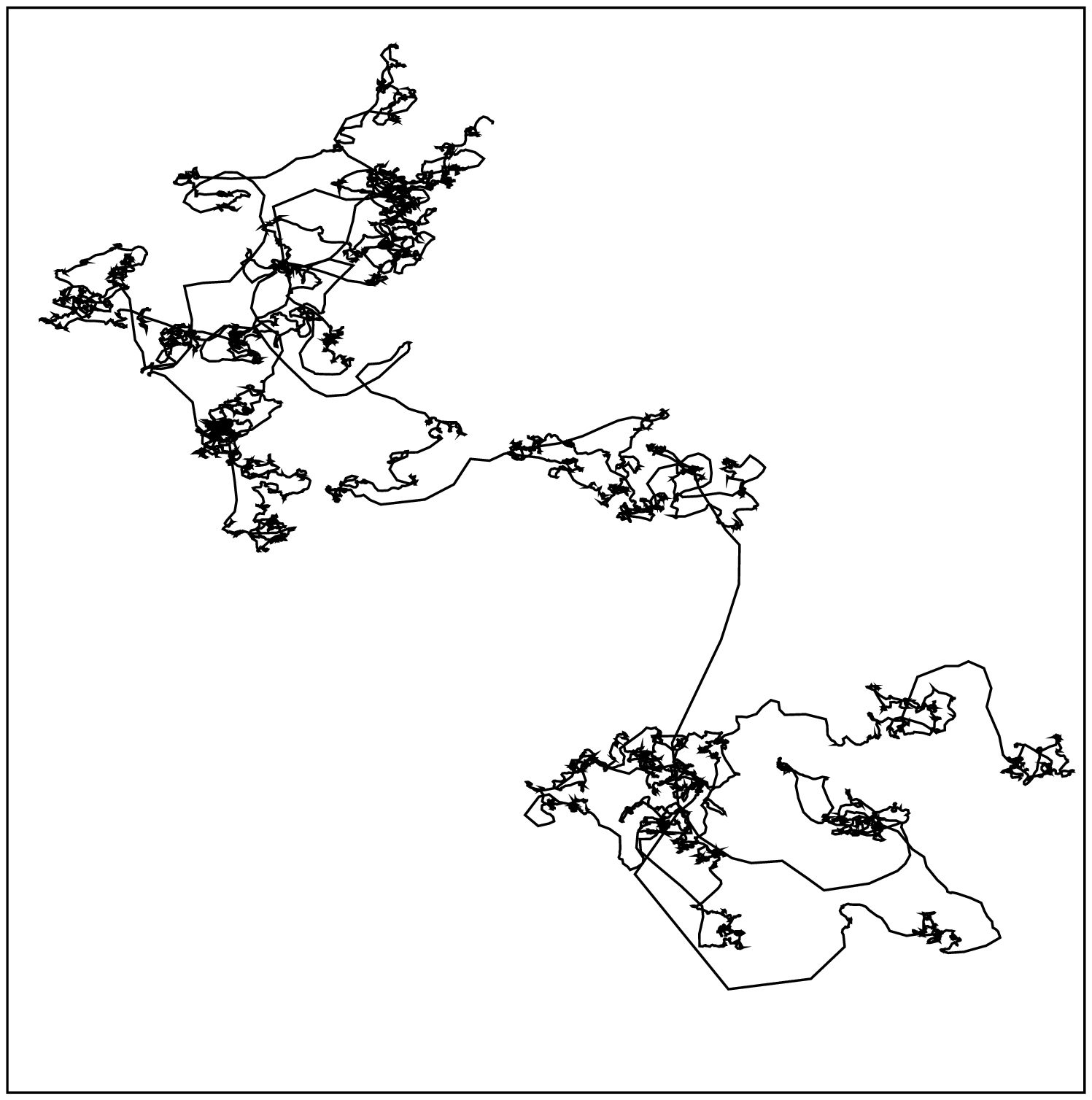}
\end{center}
\caption{Characteristic form of random walks described by 2D analogy of
model~\eqref{eq.1}. The used system parameters meet the L\'evy exponent $\alpha
= 1.6$.}
\label{F1}
\end{figure}

The purpose of the present Letter is to formulate an approach to
describing L\'evy flights and L\'evy walks using the \textit{notion
of continuous Markovian trajectories}. The key idea is to introduce
the velocity as a second variable but staying with simple Gaussian
noise. For a fixed time scale $\delta t$ we can recover the standard
behavior of L\'evy flights. However, we have full locality in the
sense that a trajectory can be determined with any desired
resolution.

In the general form the model proposed for consideration is reduced
to the class of coupled governing equations for the state vectors
$\mathbf{x}= \{x_i\}$ and $\mathbf{v} = \{v_i\}$
\begin{subequations}\label{Laneq}
\begin{align}
\label{Laneq.1}
    \frac{d\mathbf{x}}{dt} & =  \mathbf{F}(\mathbf{x}, \mathbf{v})\,,
\\
\label{Laneq.2}
    \frac{d\mathbf{v}}{dt} &= \mathbf{H}(\mathbf{x},\mathbf{v})+
    \mathbf{G}(\mathbf{x},\mathbf{v})\cdot\boldsymbol{\xi}(t)\,.
\end{align}
\end{subequations}
Here the Langevin equation~\eqref{Laneq.2} is written in the It\^o form,
$\boldsymbol{\xi}(t)=\{\xi_i(t)\}$ are the collection of mutually independent
Gaussian white noise components, the ``forces'' $\mathbf{F}(\mathbf{x},
\mathbf{v})$, and $\mathbf{H}(\mathbf{x}, \mathbf{v})$ are given functions, and
the matrix $\mathbf{G}(\mathbf{x}, \mathbf{v})$ depending on the state
variables specifies the intensity of Langevin ``forces''. In some sense we
reduce superdiffusion to a normal diffusion process expanding the phase space
where a new variable, particle velocity, is governed by the Langevin equation
with normal but multiplicative noise.

In this Letter we exemplify our procedure applying to the following 1D system
with two variables, the coordinate of random walker $x$ and its current
velocity $v$,
\begin{subequations}\label{eq.1}
\begin{align}
\label{eq.1x}
    \frac{dx}{dt}  &  = v\,,
\\
\label{eq.1v}
    \frac{dv}{dt}  &  =-\frac{(\alpha+1)}{2\tau}\,v +\frac{1}{\sqrt{\tau}}\,
    g(v)*\xi(t)\,.
\end{align}
\end{subequations}
Here $\tau$ is a certain time scale, the intensity of the Langevin random force
is given by the function
\begin{equation}\label{eq.1g}
    g(v)= \sqrt{v_a^2+v^2}\,,
\end{equation}
with the parameter $v_a$ measuring the intensity of the additive component of
Langevin force, $\xi(t)$ is white noise such that $\left<\xi(t)\xi(t')\right> =
\delta(t-t')$, and the parameter $\alpha\in(1,2)$. The Langevin
equation~\eqref{eq.1v} is written in the H\"anggi-Klimontovich form, which is
indicated by the symbol $*$. The dynamics, resulting from a 2D version of these
equations, is visualized in Fig.~\ref{F1}.

The corresponding forward Fokker-Planck equation for the distribution function
$\mathcal{P}(x-x_0,v,v_0,t)$ reads
\begin{equation}\label{FPE}
    \frac{\partial\mathcal{P}}{\partial t}
    = \frac1{2\tau} \frac{\partial}{\partial v}
    \left[g^2(v)\frac{\partial\mathcal{P}}
    {\partial v}+(\alpha+1)v\mathcal{P}\right]
     -\frac{\partial}{\partial x}\left[  v\mathcal{P}\right]\,,
\end{equation}
where the values $x_0$ and $v_0$ specify the initial position of the walker.

The distribution of the walker velocities $v_w$ is determined by the partial
distribution function
\begin{equation}\label{PV}
    P_v(v,v_0,t):= \left<\delta(v - v_w)\right>
\end{equation}
and, by virtue of \eqref{FPE}, the stationary velocity distribution
$P_v^\text{st}(v)$ meets the equality
\begin{equation*}
   g^{2}(v)\frac{\partial P_{v}^{\text{st}}}{\partial v}+
   (\alpha+1) vP_{v}^{\text{st}}=0\,,
\end{equation*}
whence we immediately get the expression
\begin{equation}\label{PVst}
    P_{v}^{\text{st}}(v)=\frac{\Gamma\left(\frac{\alpha+1}2\right)}
    {\sqrt{\pi}\,\Gamma\left(\frac{\alpha}2\right)}\,
    \frac{v_{a}^{\alpha}}{[g(v)]^{\alpha+1}}\,
\end{equation}
where $\Gamma(\ldots)$ is the Gamma-function. In addition, using the
Fokker-Planck equation for function~\eqref{PV} following directly
from \eqref{FPE} we find the expressions
\begin{subequations}\label{PVdyn}
\begin{align}
\label{PVdyn.a}
    \left<v(t)\right> & =
    v_0\exp\left[-\frac{(\alpha-1)}{2}\frac{t}{\tau}\right]\,, &
\\
\label{PVdyn.b}
    \left<v^2(t)\right>& =
    v^2_0\exp\left[(2-\alpha)\frac{t}{\tau}\right]&\text{for $v_0\gtrsim v_a$}
\end{align}
\end{subequations}
characterizing actually the relaxation of the initial velocity distribution to
its steady state form.

\begin{figure}
\begin{center}
\includegraphics[width=\columnwidth]{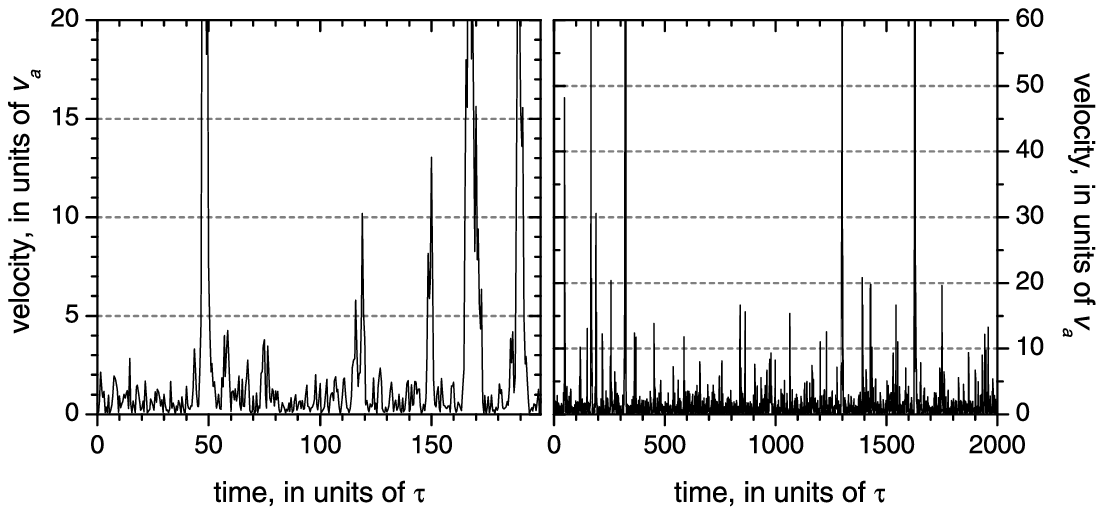}
\end{center}
\caption{Characteristic form of the time pattern $v(t)$ exhibited by
the stochastic system~\eqref{eq.1}. The individual windows depict
the patterns on various scales. In simulation the parameter $\alpha=
1.6$ was used.}
\label{F2}
\end{figure}

The found exponential decay of the first velocity moment demonstrates the fact
that the L\'evy walker ``remembers'' its velocity practically on time scales
not exceeding the value $\tau$. The exponential divergence of the second
moment~\eqref{PVdyn.b} indicates that the system relaxes to the stationary
distribution~\eqref{PVst} on time scales $t\gg\tau$. So, in some sense, the
spatial steps of duration about $\tau$ are mutually independent. In other
words, the value $\tau$ separates the time scales into two groups. On scales
less than $\tau$ the particle motion is strongly correlated and has to be
considered using both the phase variables $x$ and $v$. Thus, on a time scale
$\delta t \gg \tau$  the particle displacements are mutually independent and
the succeeding steps of the L\'evy walker form a Markovian chain, with the
particle velocity playing the role of L\'evy noise. This scenario is
exemplified in Fig.~\ref{F2} for some realization of $v(t)$ following from
equation~\eqref{eq.1v}. L\'evy flight events, i.e. the long-distance jumps of a
L\'evy walker, are due to large spikes of the time pattern $v(t)$ whose
duration is about several $\tau$. More precisely, the long-distance
displacement $\Delta x$ of a walker during a certain time interval $\delta t$
is mainly caused by the velocity spike of maximal amplitude $\vartheta$
attained during the given interval, i.e. $\Delta x\sim \vartheta\tau$.  For
 $\delta t\gg \tau$
the quantity $\{\vartheta\}$  is statistically uncorrelated during
succeeding time intervals.

Now we proceed in a two steps. First, we use this simple physical
picture to show via a combination of analytical and numerical
evidence that the distribution function $P_x(\Delta x,v_0,t):=
\left<\delta(x-x_0-\Delta x)\right>$ indeed is of
form~\eqref{LevyKernel} for $t\gg \tau$. Second, we strictly show
that the corresponding generating function fulfills
\begin{multline}\label{GFapp}
    G_x(\varkappa,t) := \Big<e^{\,\mathrm{i}(\varkappa\Delta x)/(v_a
    \tau)}\Big> \\
    {}\simeq \exp\left[ -
    \frac{\Gamma\left(\frac{2-\alpha}2\right)}{\Gamma(\alpha)\Gamma\left(\frac{\alpha}2\right)}\,
    \frac{t}{\tau}\varkappa^{\alpha} \right]\,.
\end{multline}
The latter expression is the standard generating function of L\'evy flights
with exponent $\alpha$ and matches the distribution~\eqref{LevyKernel}.

%
%

If the spikes in Fig.~\ref{F2} had the same shape and $\delta t \gg \tau$ the
normalized walker displacement $\Delta x/\vartheta$ would be a constant of the
order of $\tau$ (in the limit of large $\vartheta$ where $\Delta x$ is largely
determined by a single peak). Then $P_x(\Delta x)$ would directly follow from
the distribution of maximum velocities. To proceed we, first, make use of the
relation between the extremum statistics of Markovian processes and the first
passage time distribution \cite{Lextrema}. Namely, the probability function
$\Phi(\vartheta,v,t)$ of the random variable $\vartheta$ and the probability
$F(\vartheta,v,t)$ of passing the boundary $v=\pm \vartheta$ for the first time
at moment $t$ are related as
\begin{equation}\label{extr.1}
    \Phi(\vartheta,v,t) = -\frac{\partial }{\partial \vartheta}
    \int\limits_0^{t} dt' F(\vartheta,v,t')\,.
\end{equation}
Here $v$ is the initial position of the L\'evy walker. Then
analyzing the Laplace transform of the first passage time
distribution $F_L(\vartheta,v,s)$ we will get the conclusion that
the distribution function $\Phi(\vartheta,v,t)$ of the velocity
extrema $\vartheta$ is of the form (see the supplementary materials)
\begin{equation}\label{extr.2}
       \Phi(\vartheta,v,t) = \frac1{\bar{\vartheta}(t)}
       \phi\left[\frac{\vartheta}{\bar{\vartheta}(t)}\right ]
\end{equation}
for $v\ll\bar{\vartheta}(t)$ and $\vartheta\gtrsim \bar{\vartheta}(t)$. Here
the quantity $\bar{\vartheta}(t)=v_a(t/\tau)^{1/\alpha}$ is the velocity scale
characterizing variations of the random value $\vartheta$ and the function
$\phi(\zeta)$ possesses the asymptotics
\begin{equation}\label{extr.3}
    \phi(\zeta) = \frac{\alpha ^{2}\Gamma \left( \frac{\alpha +1}{
    2}\right) }{\sqrt{\pi }\Gamma \left( \frac{\alpha }{2}\right) }
    \frac1{\zeta^{\alpha+1}}\,.
\end{equation}

\begin{figure}
\begin{center}
\includegraphics[width=0.8\columnwidth]{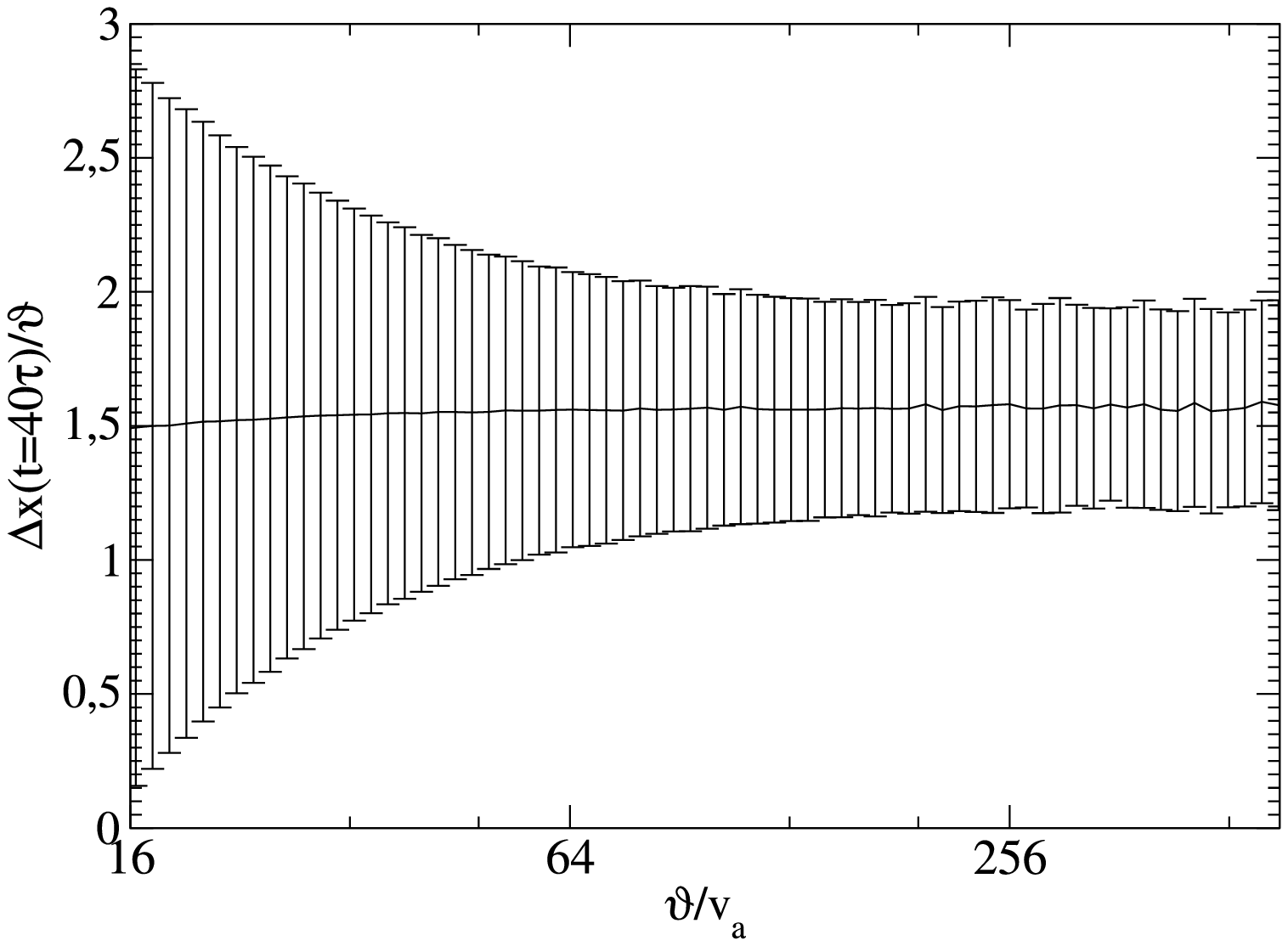}
\end{center}
\caption{The ratio $\Delta x/\vartheta$ for individual steps vs the values of
the random variable $\vartheta$. In simulation $\alpha = 1.6$ was used.}
\label{F3}
\end{figure}

Via numerical simulation we have determined the distribution of
$\Delta x / \vartheta$ for given velocity extremum $\vartheta$. The
first and second moment of this distribution is shown in
Fig.~\ref{F3}. As expected the average value of $\Delta x /
\vartheta$ indeed approaches a constant $c_\tau$ (for $\alpha = 1.6$
the value $c_\tau\approx 1.6 \tau$). However, the finite variance
shows that the velocity spikes have some distribution in their
shape. Thus a priori the distributions
$\phi\left[\frac{\vartheta}{\bar{\vartheta}(t)}\right ]$ and
$P(\Delta x)$ are not identical when replacing $\Delta x$ by $c_\tau
\vartheta$. However, since the distribution of $\Delta x /
\vartheta$ for fixed $\vartheta$ does {\it not} depend on
$\vartheta$ (for large $\vartheta$) one can directly write
\begin{multline}
P_x(\Delta x) \propto  \int d\epsilon d \vartheta q(\epsilon) \vartheta
^{-(1+\alpha)} \delta(\epsilon \vartheta + c_\tau \vartheta - \Delta x )
\\
{} \propto  \int d\epsilon q(\epsilon) [(\epsilon+c_\tau)/\Delta x]
^{(1+\alpha)} \propto \Delta x^{-(1+\alpha)}\,,
\end{multline}
where $q(\epsilon)$ is the distribution of the random variable $\epsilon :=
\Delta x / \vartheta-c_\tau$. Thus, despite the variance in peak shapes the
algebraic distribution of $\vartheta$ directly translates into an identical
distribution for $\Delta x$.

We have performed a stick derivation of formula~\eqref{GFapp} in the following
way (for details see the supplementary materials available online). The
appropriate Fokker-Planck equation should be written for the full generating
function $\mathcal{G}(\varkappa,k,t)$ for system~\eqref{eq.1} which depends on
two wave numbers, $\varkappa$ as before and $k$ related to the velocity
variations. Then the corresponding eigenvalue problem can be analyzed assuming
the wave number $\varkappa$ to be a small parameter. It turns out that the
perturbation caused by the $\varkappa$-term is singular which affects
essentially the minimal eigenvalue, making it dependent on $\varkappa$ as
$\Lambda_\text{min}\propto\varkappa^{\alpha}$. In this way
expression~\eqref{GFapp} is obtained. Furthermore the specific value of
$c_\tau$ equal to
\begin{equation}
c_\tau=\left[\frac{2\sin \left( \frac{\pi \alpha }{2}\right) \Gamma \left(
\frac{ 2-\alpha }{2}\right) }{\sqrt{\pi }\alpha \Gamma \left( \frac{\alpha
+1}{2} \right) }\right] ^{\frac{1}{\alpha }} \label{Lgood2}
\end{equation}
follows directly form the comparison of the asymptotics of $P_x(\Delta x)$
determined by \eqref{GFapp} and the asymptotics~\eqref{extr.3} of the velocity
extremum distribution. In particular, for $\alpha = 1.6$ we have $\Delta x
\approx 1.6 \vartheta\tau$ in agreement with the simulation data.

The developed model~\eqref{eq.1} actually gives us the
implementation of L\'evy flights at the ``microscopic'' level
admitting the notion of continuous trajectories. ndeed, fixing any
small duration $\delta t$ of the L\'evy walker steps we can choose
the time scale $\tau$ of model~\eqref{eq.1} such that $\delta t \gg
\tau$ and, as a result, receive the L\'evy statistics for the
corresponding spatial steps. Moreover, the found
expression~\eqref{GFapp} demonstrates the equivalence of all the
systems in asymptotic behavior for which the parameters $v_a$ and
$\tau$ are related by the expression $  v_a^\alpha \tau^{\alpha-1} =
\sigma $. In some sense, all the details of the microscopic
implementation of L\'evy flights are aggregated in two constants:
the exponent $\alpha$ and the superdiffusion coefficient $\sigma$.
In particular, the characteristic scale of the walker displacement
during time $t$ is $\bar{x}(t)\sim (\sigma t)^{1/\alpha}$.

Our approach has several immediate consequences. First of all, it
yields an easily implementable procedure for the numerical
simulation of L\'evy processes based on the simulation of the
Langevin equations (\ref{Laneq}). Second, it seems to be possible to
attack the yet unsolved problem of the formulation of accurate
boundary conditions for the generalized Fokker-Planck equations
describing L\'evy processes in finite domains and heterogeneous
media. The crucial point of our treatment is the existence of
quantities varying on three widely separated time scales
$\tilde{\delta t} \ll \tau \ll \delta t$. On time scales
$\tilde{\delta t}$ the Langevin equation is updated. In the
well-defined limit of small $\tilde{\delta t}$ the trajectory can be
constructed with arbitrary precision. Furthermore,  $\tau$ is
connected with the relaxation time of the variable $v$ and sets the
overall time scale of the model. Finally, for $\delta t$ the
variation of the position $x$ is fully Markovian and the systems
behaves according to the standard L\'evy flight scenario. A similar
approach is the treatment of the Kramers-Fokker-Planck equation
describing diffusion of particles, which is obtained from eq.
(\ref{FPE}) for the case of purely additive noise $g=const$. The
so-called Smoluchowski limit $\tau \rightarrow 0$ leads to
Einstein's diffusion equation. For equilibrium systems the
fluctuation dissipation theorem relates linear damping and purely
additive noise. The emergence of L\'evy flights, however, is related
with the presence of multiplicative noise, and, in turn, with
nonequilibrium situations.

The authors appreciate the financial support of the SFB 458 and the University
of M\"unster as well as the partial support of DFG project MA 1508/8-1 and RFBR
grants 06-01-04005, 05-01-00723, and 05-07-90248.

\begin{widetext}
\vspace{5\baselineskip}
\textbf{\large Supplementary material to paper ``Realization of L\'evy flights
as continuous processes''}
\end{widetext}

\section{Stochastic system and its governing equation}

We consider continuous 1D random walks governed by the following stochastic
differential equations of the H\"{a}nggi-Klimontovich type \cite{H1,H2,Kl}
\begin{align}
    \frac{dx}{dt}& =v\,,  \label{1} \\
    \frac{dv}{dt}& =-\frac{(\alpha +1)}{2\tau }v+\frac{1}{\sqrt{\tau }}g(v)\ast
    \xi (t)\,,  \label{2Klim}
\end{align}
where $x$ is the position of a walker, $v$ is its current velocity, $\xi (t)$
is white noise such that $\left\langle \xi (t)\xi (t^{\prime })\right\rangle
=\delta (t-t^{\prime })$, and the function
\begin{equation}
g(v)=\sqrt{v_{a}^{2}+v^{2}}  \label{4}
\end{equation}
specifies the intensity of random Langevin forces. The dimensionless
coefficient $\alpha $, the time scale $\tau $, and the characteristic velocity
$v_{a}$ quantifying the additive component of the Langevin forces are the
system parameters. The L\'{e}vy flights arise when the coefficient $\alpha $
belongs to the interval
\begin{equation}
1<\alpha <2  \label{alpha}
\end{equation}
which, thereby, is assumed to hold beforehand.

For the given system the distribution function $\mathcal{P}\left(
x-x_{0},v,v_{0},t\right) $ obeys the following Fokker-Planck equation written
in the kinetic form
\begin{equation}
    \frac{\partial \mathcal{P}}{\partial t}=\frac{1}{2\tau }\frac{\partial}
    {\partial v}\left[ g^{2}(v)\frac{\partial \mathcal{P}}{\partial v}+(\alpha
    +1)v\mathcal{P}\right] -\frac{\partial }{\partial x}\left[ v\mathcal{P}
    \right]  \label{FPKlim}
\end{equation}
subject to the initial condition
\begin{equation}
\mathcal{P}\left( x-x_{0},v,v_{0}\mathbf{,}0\right) =\delta (x-x_{0})\delta
(v-v_{0})\,,  \label{FP2}
\end{equation}
where, $x_{0}$ and $v_{0}$ are the initial position and velocity of the walker
and, an addition, the system translation invariance with respect to the
variable $x$ is taken into account.

\section{Velocity distribution\label{sec:veldist}}
\subsection{General relations}

It is the statistical properties of the walker velocity $v$ that give rise to
L\'{e}vy flights. So the present section is devoted to them individually. The
velocity distribution is given by the partial distribution function
\begin{equation}
    P_{v}(v,v_{0},t)=\int_{\mathbb{R}}dx\,\mathcal{P}\left(
    x-x_{0},v,v_{0},t\right)   \label{FP3}
\end{equation}
which, by virtue of (\ref{FPKlim}), obeys the reduced forward Fokker-Planck
equation
\begin{equation}
    2\tau \frac{\partial P_{v}}{\partial t}=\frac{\partial }{\partial v}\left[
    g^{2}(v)\frac{\partial P_{v}}{\partial v}+(\alpha +1)vP_{v}\right]
    \label{FP4f}
\end{equation}
written in the kinetic form whose right-hand side acts on the variable $v$.
Simultaneously, the function $P_{v}(v,v_{0},t)$ meets the backward
Fokker-Planck equation
\begin{equation}
    2\tau \frac{\partial P_{v}}{\partial t}=g^{2}(v_{0})\frac{\partial ^{2}P_{v}
    }{\partial v_{0}^{2}}-(\alpha -1)v_{0}\frac{\partial P_{v}}{\partial v_{0}}
    \label{FP4b}
\end{equation}
written in the \^{I}to form and acting on the variable $v_{0}$ (see, e.g.,
\cite{Gardiner}). The two equations are supplemented with the initial condition
\begin{equation}
P_{v}\left( v,v_{0},0\right) =\delta (v-v_{0})\,.  \label{FP5}
\end{equation}
In particular, as stems from (\ref{FP4f}), the stationary velocity distribution
$P_{v}^{\text{st}}(v)$ is the solution of the equation
\begin{equation}
g^{2}(v)\frac{\partial P_{v}^{\text{st}}}{\partial v}+(\alpha +1)vP_{v}^{
\text{st}}=0\,,  \label{db}
\end{equation}
which together with the normalization condition
\begin{equation}
    \int_{\mathbb{R}}dv\,P_{v}^{\text{st}}(v)=1  \label{norma}
\end{equation}
gives us the expression
\begin{equation}
P_{v}^{\text{st}}(\mathbf{v})=\frac{\Gamma \left( \frac{\alpha +1}{2}\right)
}{\sqrt{\pi }\Gamma \left( \frac{\alpha }{2}\right) }\,\frac{v_{a}^{\alpha }
}{[g(v)]^{\alpha +1}}\,,  \label{VFPStat}
\end{equation}
where $\Gamma (\ldots )$ is the Gamma function. For the exponent $\alpha $
belonging to interval~(\ref{alpha}) the first moment of the velocity $v$
converges, whereas the second one diverges what actually was the reason for
specifying the region of $\alpha $ under consideration.

\subsection{First passage time problem and extremum distribution}

In order to establish some kinematic relationship between the L\'{e}vy type
behavior exhibited by the given random walks on time scales $t\gg \tau $ and
properties of the velocity distribution we will make use of the first passage
time statistics. The probability $F(v_{0},\vartheta ,t)$ for the walker with
initial velocity $v_{0}$ such that $\left\vert v_{0}\right\vert <\vartheta $ to
gain the velocity $v=\pm \vartheta $ for the first time at the moment $t$ is
directly described by the backward Fokker-Planck equation~(\ref{FP4b}). In
particular its Laplace transform
\begin{equation*}
F_{L}(v_{0},\vartheta ,s)=\int_{0}^{\infty }dt\,e^{-st}F(v_{0},\vartheta ,t)
\end{equation*}
obeys the equation (see, e.g., \cite{Gardiner})
\begin{equation}
    2\tau sF_{L}=g^{2}(v_{0})\frac{\partial ^{2}F_{L}}{\partial v_{0}^{2}}-(\alpha
    -1)v_{0}\frac{\partial F_{L}}{\partial v_{0}}  \label{FPTP.1}
\end{equation}
subject to the boundary condition
\begin{equation}
\left. F_{L}(v_{0},\vartheta ,s)\right\vert _{v_{0}=\pm \vartheta }=1\,.
\label{FPTP.2}
\end{equation}
The introduced first passage time probability is necessary to analyze the
extremum statistics. Namely we need the probability $\Phi \left(
v_{0},\vartheta ,t\right) $ for the velocity pattern $v(t)$ originating from
the point $v_{0}\in (-\vartheta ,+\vartheta )$ to the get the extremum equal to
$\pm \vartheta $ somewhen during the time interval $t$ is related to the
probability $F(v_{0},\vartheta ,t)$ by the expression~\cite{extrema}
\begin{align}
    \Phi (v_{0},\vartheta ,t)& =-\frac{\partial }{\partial \vartheta}\int_{0}^{t}dt^{\prime
    }\,F(v_{0},\vartheta ,t^{\prime })  \label{ed:2a}
\\
\intertext{or for the Laplace transforms}
    \Phi _{L}(v_{0},\vartheta
    ,s)&=-\frac{1}{s}\frac{\partial }{\partial \vartheta }F_{L}(v_{0},\vartheta
    ,s)\,.  \label{ed:2b}
\end{align}

To examine the characteristic properties of the first passage time statistics
let us consider two limit cases, $s\rightarrow 0$ and $\vartheta \rightarrow
\infty $. Their analysis starts at the first step with the same procedure.
Namely, we assume the function $F_{L}(v_{0},\vartheta ,s)$ to be approximately
constant, $F_{L}(v_{0},\vartheta ,s)\simeq F_{0}(\vartheta ,s)$ inside some
neighborhood $\mathbb{Q}_0$ of the origin $v_{0}=0$. For $s\rightarrow 0$ it is
the domain $(-\vartheta ,\vartheta )$ itself and $F_{0}(\vartheta ,s)=1$ by
virtue of (\ref{FPTP.2}). For $\vartheta \rightarrow \infty $ the thickness of
this neighborhood is much larger then $v_{a}$ as it will be seen below. Under
such conditions equation~(\ref{FPTP.1}) can be integrated directly inside the
domain $\mathbb{Q}_0$ with respect to the formal variable $f(v_{0})=$ $\partial
F_{L}/\partial v_{0}$ using the standard parameter-variation method. In this
way taking into account that $f(0)=0$ due to the system symmetry we obtain the
expression
\begin{widetext}
\begin{align}
    \frac{\partial F_{L}(v_{0},\vartheta ,s)}{\partial v_{0}} &\simeq \frac{
    2\tau s}{v_{a}}F_{0}(\vartheta ,s)\left( \frac{v_{0}^{2}}{v_{a}^{2}}
    +1\right) ^{\frac{\alpha -1}{2}}\int_{0}^{v_{0}/v_{a}}\frac{d\xi }{\left(
    \xi ^{2}+1\right) ^{\frac{\alpha +1}{2}}}  \label{FPTP.3}
\\
\intertext{and for $\left\vert v_{0}\right\vert \gg v_{a}$}
    \frac{\partial F_{L}(v_{0},\vartheta ,s)}{\partial v_{0}} &\simeq \sqrt{\pi
    }\tau sF_{0}(\vartheta ,s)\frac{\Gamma \left( \frac{\alpha }{2}\right) }{
    \Gamma \left( \frac{\alpha +1}{2}\right) }\frac{\left\vert v_{0}\right\vert
    ^{\alpha -1}}{v_{a}^{\alpha }}\,.  \label{FPTP.5}
\end{align}
\end{widetext}

Expression~(\ref{FPTP.5}) demonstrates us that, first, the implementation of
the limit case of small values of $s$ (formally, $s\rightarrow 0$) is the
validity of the inequality
\begin{equation*}
    F_0(\vartheta,s) \gg
    v_0 \frac{\partial F_{L}(v_{0},\vartheta ,s)}{\partial v_{0}}
    \quad\Rightarrow\quad
    \tau s\frac{\left\vert v_{0}\right\vert ^{\alpha }}{v_{a}^{\alpha }}\ll
    1\,,
\end{equation*}
which can be rewritten as
\begin{align}
    \bar{\vartheta}_L(s)& :=\left( \frac1{\tau s}\right)
    ^{\frac{1}{\alpha }}v_{a}\gg |v_0|
\label{FPTP.4L}
\intertext{or converting to the time dependence}
    \bar{\vartheta}(t) & :=\left( \frac{t}{\tau }\right)
    ^{\frac{1}{\alpha }}v_{a}\gg |v_0|\,.
\label{FPTP.4}
\end{align}
So the characteristic velocity scale characterizing the first passage time
probability and aggregating its time dependence is $\bar{\vartheta}(t)$. As a
consequence, the limit of small values of $s$ is actually defined by the
inequality $|v|_0 \ll \bar{\vartheta}(t)$. Correspondingly, the limit of large
values of $\vartheta$ is implemented by the inequality $\vartheta \gg
\bar{\vartheta}(t)$ or $\vartheta \gg \bar{\vartheta}_L(s)$.

Second, for $\vartheta \gg \bar{\vartheta}_L(s)$  there is a region, namely,
$v_{a}\ll |v_{0}|\ll \bar{\vartheta}(t)$ wherein the assumption
$F_{L}(v_{0},\vartheta ,s)\simeq F_{0}(\vartheta ,s)$ holds whereas the
derivative  $\partial F_{L}/\partial v_{0}$ scales with $v_{0}$ as $\partial
F_{L}/\partial v_{0}\propto \left\vert v_{0}\right\vert ^{\alpha -1}$. This
asymptotic behavior can be obtained also by analyzing the solution of
equation~(\ref{FPTP.1}) for $\left\vert v_{0}\right\vert \gg v_{a}$ where
$g^{2}(v_{0})\simeq v_{0}^{2}$. In this case equation~(\ref{FPTP.1}) admits two
solutions of the form
\begin{gather}
\nonumber
    F_{L}(v_{0},\vartheta ,s)\propto v_{0}^{g_{1,2}}
\\
\intertext{with}
    g_{1}\simeq \alpha \quad \text{and}\quad g_{2}\simeq -\frac{2\tau s}
    {\alpha }\,.  \label{FPTP.6}
\end{gather}
The second solution is relevant to the function $F_{L}(v_{0},\vartheta ,s)$
only within the crossover from $F_{L}(v_{0},\vartheta ,s)\propto v_{0}^{\alpha
}$ to $F_{L}(v_{0},\vartheta ,s)\approx $ $F_{0}(\vartheta ,s)$ and even in
this region, i.e. $\left\vert v_{0}\right\vert \lesssim \bar{\vartheta}(t)$ the
derivative $\partial F_{L}/\partial v_{0}$ is determined by its asymptotics
$F_{L}(v_{0},\vartheta ,s)\propto v_{0}^{\alpha }$. For larger values of $v_0$,
i.e., $\left\vert v_{0}\right\vert \gg \bar{\vartheta}(t)$ the first passage
time distribution is given by the expression
\begin{equation}\label{AH.1}
    F_{L}(v_{0},\vartheta ,s)\simeq \left( \frac{\left\vert v_{0}\right\vert }{
    \vartheta }\right) ^{\alpha }
\end{equation}
taking into account the boundary condition~(\ref{FPTP.2}). So we can write
\begin{equation}\label{FPTP.7}
    \frac{\partial F_{L}(v_{0},\vartheta ,s)}{\partial v_{0}}\simeq \alpha
    \frac{\left\vert v_{0}\right\vert ^{\alpha -1}}{\vartheta ^{\alpha }}
\end{equation}
also for $\left\vert v_{0}\right\vert \lesssim \bar{\vartheta}(t)$.
Expressions~(\ref{FPTP.5}) and (\ref{FPTP.7}) describe the same asymptotic
behavior of the function $F_{L}(v_{0},\vartheta ,s)$. Thereby we can ``glue''
them together, obtaining the expression for
\begin{equation}
    F_{0}(\vartheta ,s)=\frac{\alpha \Gamma \left( \frac{\alpha +1}{2}\right) }
    {\sqrt{\pi }\Gamma \left( \frac{\alpha }{2}\right) }\frac{1}{\tau s}
    \frac{v_{a}^{\alpha }}{\vartheta ^{\alpha }}\,.  \label{FPTP.9}
\end{equation}
It should be noted that this procedure is the kernel of the singular
perturbation technique which will be also used below.
Expression~\eqref{FPTP.9} immediately gives us the desired formula for the
extremum distribution $\Phi _{L}(v_{0},\vartheta ,s)$. Namely, by virtue of
(\ref{ed:2b}), for $\left\vert v_{0}\right\vert \lesssim \bar{\vartheta}_L(s)$
we have
\begin{equation}
    \Phi _{L}(v_{0},\vartheta ,s)=\frac{\alpha ^{2}\Gamma
    \left( \frac{\alpha +1}{2}\right) }{\sqrt{\pi }
    \Gamma \left( \frac{\alpha }{2}\right) }\frac{1}
    {\tau s^{2}}\frac{v_{a}^{\alpha }}{\vartheta ^{\alpha +1}}
\label{ed:5a}
\end{equation}
and restoring the time dependence of the extremum distribution from its Laplace
transform the asymptotic behavior for $\vartheta \gg \bar{\vartheta}(t)$ we get
\begin{equation}
\Phi (v_{0},\vartheta ,t)=\frac{\alpha ^{2}\Gamma \left( \frac{\alpha +1}{2}
\right) }{\sqrt{\pi }\Gamma \left( \frac{\alpha }{2}\right) }\frac{t}{\tau }
\frac{v_{a}^{\alpha }}{\vartheta ^{\alpha +1}}\,.  \label{ed:5b}
\end{equation}

Finalizing the present subsection we draw the conclusion that for $\left\vert
v_{0}\right\vert \ll \bar{\vartheta}(t)$ the extremum distribution is described
by a certain function
\begin{equation}
    \Phi (v_{0},\vartheta ,t)=\frac{1}{\bar{\vartheta}(t)}\Phi _{0}
    \left( \frac{\vartheta }{\bar{\vartheta}(t)}\right)
\label{ad:6a}
\end{equation}
with the asymptotics
\begin{equation}
\Phi _{0}\left( \xi \right) =\frac{\alpha ^{2}\Gamma \left( \frac{\alpha +1}{
2}\right) }{\sqrt{\pi }\Gamma \left( \frac{\alpha }{2}\right) }\,\frac{1}{ \xi
^{\alpha +1}}\,.  \label{ad:6b}
\end{equation}
Here the velocity scale $\bar{\vartheta}(t)$ is given by
expression~(\ref{FPTP.4}). We remind that distribution~\eqref{ed:5b} describes
the amplitude of the velocity extrema, so, as the velocity extrema themselves
are concerned their distribution is characterized by the function
\begin{equation}\label{ppz}
 \tilde{\Phi} (v_{0},\vartheta ,t) = \frac12 \Phi (v_{0},|\vartheta|,t)
\end{equation}
because of the symmetry in the velocity fluctuations.

It should be also noted that the asymptotics $\Phi (v_{0},\vartheta ,t)\propto
\vartheta^{-(\alpha+1)}$ for $\vartheta\gg \bar{\vartheta}(t)$ could be
obtained immediately from equation~\eqref{FPTP.1}. In fact, formally assuming
$\vartheta\to\infty$ and taking into account the boundary
condition~\eqref{FPTP.2} we can represent the solution of
equation~\eqref{FPTP.1} in form~\eqref{AH.1} for $v_0\lesssim\vartheta$
because, first, $g^2(v_0)=v_0^2$ in this case and, second, the function
$F_L(v_0,\vartheta,s)$ must be decreasing with $|v_0|$. It is the only one
place where the variable $\vartheta$ enters the function
$F_L(v_0,\vartheta,s)$, thus, for $\vartheta\gg \bar{\vartheta}(t)$
\begin{equation*}
    F_L(v_0,\vartheta,s)\propto \frac{1}{\vartheta^\alpha}
\end{equation*}
and relationship~\eqref{ed:2b} directly gives rise to
\begin{equation*}
    \Phi (v_{0},\vartheta ,t)\propto \frac{1}{\vartheta^{\alpha+1}}\,.
\end{equation*}

\section{Generating function\label{sec:genfun}}
\subsection{General relations}

To analyze the given stochastic process the generating function
\begin{equation}
\mathcal{G}(k,\varkappa ,t)=\left\langle \exp \left\{ \frac{i}{v_{a}\tau }
\left[ \tau vk+\left( x-x_{0}\right) \varkappa \right] \right\} \right\rangle
\label{GenF}
\end{equation}
is introduced. As follows from the Fokker-Planck equation~(\ref{FPKlim}) it
obeys the governing equation
\begin{equation}
2\tau \frac{\partial \mathcal{G}}{\partial t}=\frac{\partial }{\partial k}
\left( k^{2}\frac{\partial \mathcal{G}}{\partial k}\right) +\left[ 2\varkappa
-(\alpha +1)k\right] \frac{\partial \mathcal{G}}{\partial k}-k^{2} \mathcal{G}
\label{FPGenF}
\end{equation}
subject to the initial condition
\begin{equation}
\mathcal{G}(k,\varkappa ,0)=\exp \left\{ \frac{i}{v_{a}}v_{0}k\right\} \,.
\label{GenFICond}
\end{equation}
At the origin $k=0$ and $\varkappa =0$ function~(\ref{GenF}) meets also the
identity
\begin{equation}
\mathcal{G}(0,0,t)=1  \label{GenFNorm}
\end{equation}%
which follows directly from the meaning of probability. In deriving
equation~(\ref{FPGenF}) the following relationships between the operators
acting in the spaces $\left\{ x,v\right\} $ and $\left\{ \varkappa ,k\right\} $
\begin{equation*}
\frac{\partial }{\partial x}\leftrightarrow -\frac{i}{v_{a}\tau }\varkappa
\,,\quad \frac{\partial }{\partial v}\leftrightarrow -\frac{i}{v_{c}} k\,,\quad
v=-iv_{a}\frac{\partial }{\partial k}
\end{equation*}
as well as the commutation rule
\begin{equation*}
\frac{\partial }{\partial k}k-k\frac{\partial }{\partial k}=1
\end{equation*}
have been used.

The argument $\varkappa $ enters equation (\ref{FPGenF}) as a parameter; the
given equation does not contain any differential operator acting upon the
function $\mathcal{G}(k,\varkappa ,t)$ via the argument $\varkappa $. This
property enables us to pose a question about the spectrum of
equation~(\ref{FPGenF}), where the variable $\varkappa $ plays the role of a
parameter. The desired eigenfunctions and their eigenvalues
\begin{equation}
\left\{ \Psi _{\Lambda }\left( k|\varkappa \right) \right\} ,\qquad \left\{
\Lambda \left( \varkappa \right) \right\}  \label{eigenFV}
\end{equation}
obey the equation
\begin{equation}
-2\Lambda \Psi _{\Lambda }=\frac{d}{dk}\left( k^{2}\frac{d\Psi _{\Lambda }}{
dk}\right) +\left[ 2\varkappa -(\alpha +1)k\right] \frac{d\Psi _{\Lambda }}{
dk}-k^{2}\Psi _{\Lambda }\,.  \label{Eigen1}
\end{equation}
In deriving equation (\ref{Eigen1}) the time dependence $\exp (-\Lambda t/\tau
)$ corresponding to eigenfunctions~(\ref{eigenFV}) has been assumed.

In these terms the solution of equation~(\ref{FPGenF}) is reduced to the series
\begin{equation}
\mathcal{G}(k,\varkappa ,t)=\sum_{\Lambda }f_{\Lambda }\left( \varkappa
,|v_{0}\right) \Psi _{\Lambda }\left( k|\varkappa \right) \exp \left\{ -\Lambda
\left( \varkappa \right) \frac{t}{\tau }\right\}  \label{Series}
\end{equation}
whose the coefficients $\left\{ f\left( \varkappa ,|v_{0}\right) \right\} $
meet the equality
\begin{equation}
\sum_{\Lambda }f_{\Lambda }\left( \varkappa ,|v_{0}\right) \Psi _{\Lambda
}\left( k|\varkappa \right) =\exp \left\{ \frac{i}{v_{a}}v_{0}k\right\}
\label{coeff}
\end{equation}
steaming from the initial condition~(\ref{GenFICond}). In agreement with the
results to be obtained, the spectrum of the Fokker-Planck
equation~(\ref{FPGenF}) is bounded from below by a nondegenerate minimal
eigenvalue $\Lambda _{\text{min}}\left( \varkappa \right)\geq0 $ whereas the
other eigenvalues are separated from it by a final gap of order unity.

So, as time goes on and the inequality $t\gg \tau $ holds, the term
corresponding to the minimal eigenvalue will be dominant and sum~(\ref{Series})
is reduced to
\begin{equation}
\mathcal{G}(k,\varkappa ,t)=f_{\text{min}}\left( \varkappa ,|v_{0}\right) \Psi
_{\text{min}}\left( k|\varkappa \right) \exp \left\{ -\Lambda _{\text{
min}}\left( \varkappa \right) \frac{t}{\tau }\right\}  \label{Series1}
\end{equation}
on large time scales. Here $\Psi _{\text{min}}\left( k|\varkappa \right)$ is
the eigenfunction of the eigenvalue $\Lambda _{\text{min}}\left( \varkappa
\right) $.

Whence several consequences follow. First, the identity~(\ref{GenFNorm}) holds
at any time moment, thereby
\begin{equation}
\Lambda _{\text{min}}\left( 0\right) =0\,.  \label{Series2}
\end{equation}
Second, in the limit case $t\gg \tau $ the system has to \textquotedblleft
forget\textquotedblright\ the value $v_{0}$ of initial velocity, so the
coefficient $f_{\text{min}}\left( \varkappa \right) $ does not depend on the
argument $v_{0}$ and, therefore, can be aggregated into the function $\Psi
_{\text{min}}\left( k|\varkappa \right) $. In this way the initial condition
expansion~(\ref{coeff}) reads
\begin{equation}
\Psi _{\text{min}}\left( k|\varkappa \right) +\sum_{\Lambda >\Lambda _{\text{
min}}}f_{\Lambda }\left( \varkappa ,|v_{0}\right) \Psi _{\Lambda }\left(
k|\varkappa \right) =\exp \left\{ \frac{i}{v_{a}}v_{0}k\right\} \label{coeff2}
\end{equation}
for any $v_{0}$.

The terms in sum~\eqref{coeff2} with $\Lambda
>\Lambda_\text{min}$ determine the dependence of the generating function $\mathcal{G}(k,\varkappa
,t)$ on the initial velocity $v_0$, so, the corresponding coefficients
$f_{\Lambda }\left( \varkappa ,|v_{0}\right)$ must depend on $v_0$. Finding the
first derivative of both the sides of this equality with respect to $v_{0}$ we
have
\begin{equation*}
\sum_{\Lambda >\Lambda _{\text{min}}}\frac{\partial }{\partial v_{0}}
f_{\Lambda }\left( \varkappa ,|v_{0}\right) \Psi _{\Lambda }\left( k|\varkappa
\right) =\frac{i}{v_{a}}k\exp \left\{ \frac{i}{v_{a}} v_{0}k\right\}\,.
\end{equation*}
Whence it follows that, third, the eigenfunctions $\Psi _{\Lambda }\left(
k|\varkappa \right) $ for $\Lambda >\Lambda _{\text{min}}$ must exhibit the
asymptotic behavior $\Psi _{\Lambda }\left( k|\varkappa \right) \rightarrow 0 $
as $k\rightarrow 0$ because of their linear independence. Fourth, setting $k=0$
in expression~(\ref{coeff2}) we get the conclusion that the eigenfunction $\Psi
_{\text{min}}\left( k|\varkappa \right) $ has to meet the normalization
condition
\begin{equation}
\Psi _{\text{min}}\left( 0|\varkappa \right) =1\quad \text{at}\quad k=0\,.
\label{PsiNormal}
\end{equation}

Summarizing the aforementioned we see that on large time scales $t\gg \tau $
the desired asymptotic behavior of the given system is described by the
generating function
\begin{align}
    \mathcal{G}(k,\varkappa ,t)& =\Psi _{\text{min}}\left( k|\varkappa \right) \exp
    \left\{ -\Lambda _{\text{min}}\left( \varkappa \right) \frac{t}{\tau }
    \right\} \,,
\label{IamH1}
\\
\intertext{and by virtue of (\ref{PsiNormal})}
    \mathcal{G}(0,\varkappa ,t)&
    =\exp \left\{ -\Lambda _{\text{min}}\left( \boldsymbol{\varkappa }\right)
    \frac{t}{\tau }\right\} \,.  \label{IamH2}
\end{align}
In what follows the calculation of the eigenvalue $\Lambda _{\text{min}%
}\left( \varkappa \right) $ will be the main goal.

The given random walks should exhibit the L\'{e}vy flight behavior on large
spatial and temporal scales, i.e. $x\gg v_{a}\tau $ and $t\gg \tau $. It allows
us to confine our analysis to the limit of small values of $\varkappa $, i.e.
assume that $\left\vert \varkappa \right\vert \ll 1$, where also the eigenvalue
$\Lambda_\text{min}(\varkappa)\ll 1$. In this case the spectrum of
equation~(\ref{FPGenF}) may be studied using perturbation technique, where the
term
\begin{equation}
\widehat{V}_{\varkappa }\Psi =2\varkappa \frac{d\Psi }{dk}  \label{perturb}
\end{equation}
plays the role of perturbation.

\subsection{Zero-th approximation. Spectral properties of the velocity
distribution}

The zero-th approximation of \eqref{FPGenF} in perturbation (\ref{perturb})
matches the case $\varkappa =0$, where the generating function~(\ref{GenF})
actually describes the velocity distribution~(\ref{FP3}). Setting $\varkappa
=0$ reduces the eigenvalue equation~(\ref{Eigen1}) to the following
\begin{equation}
    -2\lambda \Phi _{\lambda }=\frac{d}{dk}\left( k^{2}\frac{d\Phi _{\lambda }}{
    dk}\right) -(\alpha +1)k\frac{d\Phi _{\lambda }}{dk}-k^{2}\Phi _{\lambda }\,,
\label{Eigen0}
\end{equation}
where
\begin{equation}
    \Phi _{\lambda }(k)=\Psi _{\Lambda }\left( k|0\right) \quad \text{and}\quad
    \lambda =\Lambda \left( 0\right) \,.
\label{eigenFV0}
\end{equation}
Having in mind different goals we consider the conversion of
equation~(\ref{Eigen0}) under the replacement
\begin{equation}
\Phi _{\lambda }(k)=\left\vert k\right\vert ^{\beta _{i}}\phi _{\lambda
,i}\left( k\right)  \label{evProb1}
\end{equation}
for two values of the exponent $\beta _{i}$.

First, for $\beta _{1}=(\alpha +1)/2$ equation~(\ref{Eigen0}) is converted into
\begin{equation}
2\lambda \phi _{\lambda ,1}=-\frac{d}{dk}\left( k^{2}\frac{d\phi _{\lambda
,1}}{dk}\right) +\left[ k^{2}+\frac{1}{4}(\alpha ^{2}-1)\right] \phi _{\lambda
,1}\,.  \label{eqn1}
\end{equation}
The operator on the right-hand side of equation~(\ref{eqn1}) is Hermitian
within the standard definition of scalar product. So all the eigenvalues
$\lambda $ are real numbers and the corresponding eigenfunctions form a basis.
It should be noted that the given conclusion coincides with the well known
property of the Fokker-Planck equations with the detailed balance
\cite{Risken}. In addition the eigenfunctions $\phi _{\lambda ,1}\left(
k\right) $ can be chosen so that the identity
\begin{equation}
\int_{\mathbb{R}}dk\,\phi _{\lambda ,1}^{\ast }\left( k\right) \phi _{\lambda
^{\prime },1}\left( k\right) =\delta _{\lambda \lambda ^{\prime }}
\label{orthog}
\end{equation}
holds for all of them except for the eigenfunction $\phi _{\text{min}}\left(
k\right) $ corresponding to the minimal eigenvalue $\lambda
_{\text{min}}=\Lambda _{\text{min}}\left( 0\right) =0$ by virtue of
(\ref{Series2}). We note that the latter eigenfunction describes the stationary
velocity distribution~(\ref{VFPStat}) and its normalization is determined by
condition~(\ref{PsiNormal}). Treating the eigenfunction $\Phi _{\text{min}
}\left( k\right) $ individually releases the remainders from the necessity to
take a nonzero value at the origin $k=0$ and, thereby, enables the
eigenfunction problem~(\ref{eqn1}) to be considered within $L^{2}$-space.

Second, for $\beta _{2}=\alpha /2$ equation~(\ref{Eigen0}) is reduced to the
modified Bessel differential equation
\begin{equation}
k^{2}\frac{d^{2}\phi _{\lambda ,2}}{dk^{2}}+k\frac{d\phi _{\lambda ,2}}{dk}-
\left[ k^{2}+\frac{1}{4}\alpha ^{2}-2\lambda \right] \phi _{\lambda ,2}=0\,.
\label{bessel}
\end{equation}
Since the desired eigenfunctions should decrease as $k\rightarrow \infty $ the
solution of equation~(\ref{bessel}) is given by the modified Bessel function of
the second kind
\begin{equation}
\phi _{\lambda ,2}(k)\propto K_{\nu }(\left\vert k\right\vert )  \label{Knu}
\end{equation}
with the order $\nu =\sqrt{\frac{1}{4}\alpha ^{2}-2\lambda }$ because
\begin{equation*}
K_{\nu }(\left\vert k\right\vert )\sim \sqrt{\frac{\pi }{2\left\vert
k\right\vert }}e^{-\left\vert k\right\vert }\quad \text{as}\quad k\rightarrow
\infty
\end{equation*}
for any value of the parameter $\nu $ \cite{specfun}.

Whence it follows that there are no eigenfunctions with $\lambda
<\frac{1}{8}\alpha ^{2}$ and $\lambda \neq 0$. Indeed, when $\lambda <0$ the
function
\begin{equation*}
\Phi (k):=\left\vert k\right\vert ^{\frac{1}{2}\alpha }K_{\nu }(\left\vert
k\right\vert )\propto \left\vert k\right\vert ^{-(\nu -\frac{1}{2}\alpha
)}\quad \text{for}\quad \left\vert k\right\vert \ll 1
\end{equation*}
diverges as $k\rightarrow 0$. In the region $0<\lambda <\frac{1}{8}\alpha ^{2}$
the corresponding eigenfunctions
\begin{equation*}
\phi _{\lambda ,1}(k)=\phi _{\lambda ,3}(k)\left\vert k\right\vert ^{\beta
_{2}-\beta _{1}}\propto \left\vert k\right\vert ^{-\frac{1}{2}}K_{\nu
}(\left\vert k\right\vert )
\end{equation*}
would give rise to a strong divergency in the normalization
condition~(\ref{orthog}). When $\lambda >\frac{1}{8}\alpha ^{2}$ the solution
of equation~(\ref{bessel}) is described by the modified Bessel functions of
pure imaginary order which exhibit strongly oscillatory behavior as
$\varkappa\to0$ and describe the continuous spectrum of the Fokker-Planck
equation~(\ref{FPGenF}) for $ \varkappa =0$. Due to result~(\ref{Knu}) the
eigenfunction $\Phi _{\text{min} }\left( k\right) $ corresponding to the
eigenvalue $\lambda =0$ and meeting the normalization
condition~(\ref{PsiNormal}) is of the form
\begin{multline}
\Phi _{\text{min}}\left( k\right) =\frac{2^{\frac{2-\alpha }{2}}}{\Gamma (
\frac{\alpha }{2})}k^{\frac{\alpha }{2}}K_{\frac{\alpha }{2}}(\left\vert
k\right\vert )
\\
{}=1-\left( \frac{\left\vert k\right\vert }{2}\right) ^{\alpha } \frac{\Gamma
\left( \frac{2-\alpha }{2}\right) }{\Gamma \left( \frac{ 3+\alpha }{2}\right)
}+O(k^{2})\,. \label{VDzeroeigen}
\end{multline}
In deriving expression~(\ref{VDzeroeigen}) the following expansion of the
function $K_{\nu }(k)$ has been used
\begin{equation}
K_{\nu }(\left\vert k\right\vert )=\frac{\Gamma (\nu )}{2^{1-\nu }\left\vert
k\right\vert ^{\nu }}\left[ 1-\left( \frac{\left\vert k\right\vert }{2} \right)
^{2\nu }\frac{\Gamma (1-\nu )}{\Gamma (1+\nu )}+O(k^{2})\right] \label{Knuexp}
\end{equation}
which is justified for the order $0<\nu <1$ (see, e.g., Ref.~\cite{specfun}).
The latter inequality holds due to the adopted assumption~(\ref{alpha}) about
the possible values of the parameter $\alpha $.

Expression~(\ref{VDzeroeigen}) finalizes the analysis of the zero-th
approximation. Summarizing the aforementioned we draw the conclusion that at
$\varkappa =0$ the spectrum of the Fokker-Planck equation~(\ref{FPGenF}) for
the generating function~(\ref{GenF}) does contain zero eigenvalue $\Lambda _{
\text{min}}\left( 0\right) =0$ corresponding to
eigenfunction~(\ref{VDzeroeigen}) which is separated from higher eigenvalues by
a gap equal to $\alpha ^{2}/8$ (in units of $\tau $). We note that the given
statement is in agreement with the conclusion about the spectrum properties for
a similar stochastic process with multiplicative noise~\cite{MN1,MN2,MN3}.

\subsection{The eigenvalue $\Lambda _{\text{min}}(\varkappa)$ for
$\left\vert\varkappa\right\vert \ll 1 $. Singular perturbation technique}

When $\varkappa \neq 0$ the perturbation term (\ref{perturb}) mixes the
eigenfunctions of zero-th approximation and, as a result, the eigenfunctions
$\Phi _{\lambda }(k)$ with $\lambda >0$ contribute also to the eigenfunction
$\Psi _{\text{min}}\left( k|\varkappa \right) $. However, because their
eigenvalues are about unity or larger, $\lambda \gtrsim 1$, the perturbation
can be significant only in the domain $\left\vert k\right\vert \lesssim
\left\vert \varkappa \right\vert $. Outside this domain the perturbation is not
essential and the eigenfunction $\Psi _{\text{min} }\left( k|\varkappa \right)
$ practically coincides with its its zero-th approximation $\Phi
_{\text{min}}\left( k\right) $. So in the case when $ \left\vert
\boldsymbol{\varkappa }\right\vert \ll 1$ there should be an interval
$\left\vert \varkappa \right\vert \ll \left\vert k\right\vert \ll 1$ where, on
one hand, the eigenfunction $\Psi _{\text{min}}\left( k|\varkappa \right) $ can
be already approximated by $\Phi _{\text{min}}\left( k\right) $ and, on the
other hand, the expansion~(\ref{VDzeroeigen}) still holds, in particular, $\Psi
_{\text{min}}\left( k|\varkappa \right) \approx 1$ in this region. Leaping
ahead, we note that $\Lambda _{\text{min}}\sim \left\vert \varkappa \right\vert
^{\alpha }$ so inside the subinterval $\left\vert \varkappa \right\vert \ll
\left\vert k\right\vert \ll \left\vert \varkappa \right\vert ^{\frac{\alpha
}{2}}$ the last term on the right-hand side of equation~(\ref{Eigen1}) is also
ignorable in comparison with its left-hand side. Under these conditions the
eigenvalue equation~(\ref{Eigen1}) is reduced to the following
\begin{equation}
2\Lambda _{\text{min}}=\frac{d}{dk}\left( k^{2}\frac{d\psi }{dk}\right) +
\left[ 2\varkappa -(\alpha +1)k\right] \frac{d\psi }{dk}  \label{spt2}
\end{equation}
for the function $\psi \left( k|\varkappa \right) =1-\Psi _{\text{min} }\left(
k|\varkappa \right) $. In the given case the singular perturbation technique is
implemented within the replacement $k=\zeta \varkappa $ converting
equation~(\ref{spt2}) into one of the form
\begin{equation}
    2\Lambda _{\text{min}}=\frac{d}{d\zeta }\left( \zeta ^{2}\frac{d\psi }{dk}
    \right) +\left[ 2-(\alpha +1)\zeta \right] \frac{d\psi }{d\zeta }
\label{spt2a}
\end{equation}
subject to the effective ``boundary'' conditions by virtue of
(\ref{VDzeroeigen})
\begin{equation}
\psi \left( \zeta |\varkappa \right) \sim \left\vert \zeta \right\vert ^{\alpha
}\left( \frac{\left\vert \varkappa \right\vert }{2}\right) ^{\alpha
}\frac{\Gamma \left( \frac{2-\alpha }{2}\right) }{\Gamma \left( \frac{ 3+\alpha
}{2}\right) }\quad \text{as}\quad \zeta \rightarrow \pm \infty \,.
\label{spt2b}
\end{equation}
In some sense the condition~(\ref{spt2b}) ``glues'' the asymptotic behavior of
the eigenfunction $\Psi _{\text{min}}\left( k|\varkappa \right) $ resulting
from its properties for sufficiently large values of \thinspace \thinspace $k$
together with one stemming from small values of $k$, in this case, specified by
the solution of equation~(\ref{spt2b}). Exactly such a procedure is the essence
of the singular perturbation technique.

Equation~(\ref{spt2a}) with respect to the variable $d\psi /d\zeta $ can be
solved directly using the standard parameter-variation method. In this way we
get for $\zeta <0$
\begin{widetext}
\begin{align}
\frac{d\psi }{d\zeta } & = \left\vert \zeta \right\vert ^{\alpha -1}\left[ \exp
\left( \frac{2}{\zeta }\right) C_{-\infty }+2^{1-\alpha }\Lambda _{
\text{min}}\int\limits_{2/\zeta }^{0}\xi ^{\alpha -1}\exp \left( \frac{2}{
\zeta }-\xi \right) d\xi \right] \label{1D:2a}
\\
\intertext{and for $\zeta>0$}
\frac{d\psi }{d\zeta } &=\left\vert \zeta \right\vert ^{\alpha -1}\exp \left(
\frac{2}{\zeta }\right) \left[ C_{+\infty }-2^{1-\alpha }\Lambda _{
\text{min}}\int\limits_{0}^{2/\zeta }\xi ^{\alpha -1}\exp \left( -\xi \right)
d\xi \right] \,, \label{1D:2b}
\end{align}
\end{widetext}
where the constants $C_{\pm \infty }$ specify the asymptotic behavior of the
derivative
\begin{equation*}
\frac{d\psi }{d\zeta }\sim \left\vert \zeta \right\vert ^{\alpha -1}C_{\pm
\infty }\quad \text{as}\quad \zeta \rightarrow \pm \infty
\end{equation*}
and according to condition~(\ref{spt2b})
\begin{equation}
C_{+\infty }=-C_{-\infty }=\alpha \left( \frac{\left\vert \varkappa \right\vert
}{2}\right) ^{\alpha }\frac{\Gamma \left( \frac{2-\alpha }{2} \right) }{\Gamma
\left( \frac{3+\alpha }{2}\right) }\,.  \label{1D:2c}
\end{equation}
Expression (\ref{1D:2b}) diverges as $\zeta \rightarrow 0$ unless the equality
\begin{equation*}
\frac{\Gamma (1-\nu )}{\Gamma (\nu )}-\Lambda _{\text{min}
}\int\limits_{0}^{\infty }\exp \left( -\xi \right) \xi ^{\alpha -1}d\xi =0
\end{equation*}
holds, whence we find the desired expression for the eigenvalue $\Lambda _{
\text{min}}$
\begin{equation}
\Lambda _{\text{min}}=\frac{\Gamma \left( \frac{2-\alpha }{2}\right) }{ \Gamma
\left( \alpha \right) \Gamma \left( \frac{\alpha }{2}\right) } \left\vert
\varkappa \right\vert ^{\alpha }\,.  \label{1D:3}
\end{equation}

Expression~(\ref{1D:3}) finalizes the analysis of the generating
function~(\ref{GenF}). In particular, together with expression~(\ref{IamH2}) it
gives the desired formula for the generating function
\begin{multline}
\mathcal{G}(0,\varkappa ,t)=\left\langle \exp \left\{ i\frac{\left(
x-x_{0}\right) \varkappa }{v_{a}\tau }\right\} \right\rangle
\\
{} =\exp \left\{ - \frac{\Gamma \left( \frac{2-\alpha }{2}\right) }{\Gamma
\left( \alpha \right) \Gamma \left( \frac{\alpha }{2}\right) }\left\vert
\varkappa \right\vert ^{\alpha }\frac{t}{\tau }\right\}   \label{1D:F1}
\end{multline}
demonstrating the fact that the given random walks exhibit L\'{e}vy flight
statistics on time scales $t\gg \tau $. Expression~(\ref{1D:F1}) in turn gives
us the asymptotics of the $x$-distribution function
\begin{equation*}
P_{x}(x-x_{0},v_{0},t)=\int_{\mathbb{R}}dv\,\mathcal{P}\left(
x-x_{0},v,v_{0},t\right)
\end{equation*}
for $\left\vert x-x_{0}\right\vert \gg \bar{x}(t)$ in the form
\begin{equation}
P_{x}(x-x_{0},t)=\frac{\sin \left( \frac{\pi \alpha }{2}\right) \alpha \Gamma
\left( \frac{2-\alpha }{2}\right) }{\pi \Gamma \left( \frac{\alpha }{2}\right)
}\,\frac{\bar{x}^{\alpha }(t)}{\left\vert x-x_{0}\right\vert ^{\alpha +1}}
\label{1D:F2}
\end{equation}
where the length
\begin{equation}
\bar{x}(t)=\left( \sigma t\right) ^{\frac{1}{\alpha }} \label{1D:F3}
\end{equation}
with $\sigma =v_{a}^{\alpha }\tau ^{\alpha -1}$ specifies the characteristic
scales of the walker displacement during the time interval $t$.

\section{The L\'{e}vy flight behavior and the extremum statistics of the
walker velocities}

Comparing expressions (\ref{ppz}) and (\ref{1D:F2}) describing the asymptotic
behavior of the given random walks with respect to the walker displacement
$x-x_{0}$ and its velocity extrema $\vartheta $ we get the relationship between
their characteristic scales
\begin{equation}
\bar{x}(t)=\bar{\vartheta}(t)\tau   \label{good1}
\end{equation}
and the asymptotic equivalence within the replacement $(x-x_{0})=$ $\vartheta
T$, where
\begin{equation}
T=\left[\frac{2\sin \left( \frac{\pi \alpha }{2}\right) \Gamma \left( \frac{
2-\alpha }{2}\right) }{\sqrt{\pi }\alpha \Gamma \left( \frac{\alpha +1}{2}
\right) }\right] ^{\frac{1}{\alpha }}\tau \,.  \label{good2}
\end{equation}
The obtained expressions allow us to consider the long distance displacements
of the walker within the time interval $t$ to be implemented during one spike
of duration $\tau $ in the pattern $v(t)$ that has the maximal amplitude. In
particular, for $\alpha = 1.6$ the ratio $T/\tau \simeq 1.6$

\end{document}